\documentclass[10pt, a4paper, nosuperscriptaddress, twocolumn, showpacs, titlepage, showkeys]{revtex4-1}
\usepackage{times}
\usepackage{amsmath}
\usepackage{amssymb}
\usepackage{graphicx}
\usepackage{subfigure}
\usepackage{booktabs}
\usepackage{rotating}
\usepackage{longtable}
\usepackage{bm}
\makeatletter

\newcommand{\Rmnum}[1]{\expandafter\@slowromancap\romannumeral #1@}
\makeatother

\begin{document}
\title{Multipartite entanglement of fermionic systems in noninertial frames revisited}
\author{Wen-Chao Qiang$^1$}
\email[Corresponding author: ]{qwcqj@163.com (Wen-Chao Qiang).}
%\thanks{Corresponding author:\\
%E-mail address: qwcqj@163.com (Wen-Chao Qiang).}
\author{Guo-Hua Sun$^2$}
\email{sunghdb@yahoo.com (G.H. Sun).}
%\thanks{E-mail address: sunghdb@yahoo.com (G.H. Sun).}
\author{Oscar Camacho-Nieto$^3$}
\email{ocamacho@ipn.mx (O. Camacho-Nieto).}
%\thanks{E-mail address: ocamacho@ipn.mx (O. Camacho-Nieto)}
\author{Shi-Hai Dong$^3$}
\email{dongsh2@yahoo.com (S.H. Dong).}
%\thanks{E-mail address: dongsh2@yahoo.com (S.H. Dong)}
\affiliation{$^1$ Faculty of Science, Xi'an University of Architecture and Technology, Xi'an, 710055, China\\
$^2$ Catedr\'{a}tica CONACYT, CIC-IPN, Av. Juan de Dios B\'{a}tiz, Esq. Miguel Oth\'{o}n de Mendiz\'{a}bal, Col. Nueva Industrial Vallejo, Delegaci\'{o}n Gustavo A. Madero, C.P 07738, Mexico\\
$^3$ Laboratorio de Informaci\'{o}n Cu\'{a}ntica, CIDETEC, Instituto Polit\'{e}cnico Nacional, UPALM, CDMX 07700, Mexico}
\begin{abstract}
In this work we have revisited a few principal formulae about one-tangle of multipartite entanglement of fermionic systems in noninertial frames calculated in the paper [Phys. Rev. A 83, 022314(2011)] and given their correct expressions.
\end{abstract}
\maketitle
Studying entanglements of quantum systems in a noninertial frame is an important field of quantum information theory. In a seminal work, P. M. Alsing \textit{et al} analyzed the entanglement between two modes of a free Dirac field regarded as two relatively accelerated parties \cite{PRA_74_(206)_032326}. Wang and Jing extended Alsing's study to three observers. They assumed Alice, Bob and Charlie initially share a Greenberger-Horne-Zeilinger (GHZ) state, and then let Alice keep stationary while Bob and Charlie move with uniform acceleration. They investigated this tripartite entanglement of the fermionic systems in noninertial frames using $\pi$-tangle \cite{PRA_83_(2011)_022314}. The authors of \cite{PRA_83_(2011)_022314} considered the fact that "Bob and Charlie's subsystems are symmetrical in this case". This implies that Eq.(15) in \cite{PRA_83_(2011)_022314} keeps unchanged if we exchange acceleration parameters $r_b$ and $r_c$ each other, Eqs.(16) and (17) in \cite{PRA_83_(2011)_022314} should be exchanged each other. The symmetrical properties of Eqs.(15), (16) and (17) between $r_b$ and $r_c$ are not appeared explicitly. To illustrate their symmetries, we can rewrite them in the following form
\begin{subequations}
\begin{eqnarray}
  \mathcal{N}_{A(B_IC_I)} &=&\frac{1}{2} [\cos r_b \cos r_c-\sin ^2r_b \sin ^2r_c\nonumber\\
  &+&\sqrt{\sin ^4r_b \sin ^4r_c+\cos ^2r_b \cos ^2r_c}], \label{NABICI}\\
  \mathcal{N}_{B_I(AC_I)} &=& \frac{1}{2}[\cos r_b \cos r_c-\sin ^2r_b\cos ^2r_c\nonumber\\
  &+&\cos r_c \sqrt{\sin ^4r_b \cos ^2r_c+\cos ^2r_b}], \label{NBIACI}\\
  \mathcal{N}_{C_I(AB_I)} &=& \frac{1}{2} [\cos r_b \cos r_c-\cos ^2r_b \sin ^2r_c\nonumber\\
  &+&\cos r_b \sqrt{\cos ^2r_b \sin ^4r_c+\cos ^2r_c}].\label{NCIABI}
\end{eqnarray}
\end{subequations}
%Unfortunately, the authors did not apply this symmetrical property of the system to check basic equations (15)-(17) of \cite{PRA_83_(2011)_022314}. In fact, it is easy to find Eqs.(15)-(17) in \cite{PRA_83_(2011)_022314} are not symmetrical for $r_b$ and $r_c$.
  As key formulae and main calculating results, equations (15)-(17) in \cite{PRA_83_(2011)_022314} are incorrect, all successive calculations related to them are incorrect accordingly. Considering this paper was cited more than thirty times and had a certain impact on quantum information theory since its publication in 2011, we shall recalculate three one-tangles and give correct formulae for the negativities $\mathcal{N}_{A(B_IC_I)}, \mathcal{N}_{B_I(AC_I)}$ and $\mathcal{N}_{C_I(AB_I)}$ in this paper.

\begin{figure}[htbp]
\includegraphics[width=6cm]{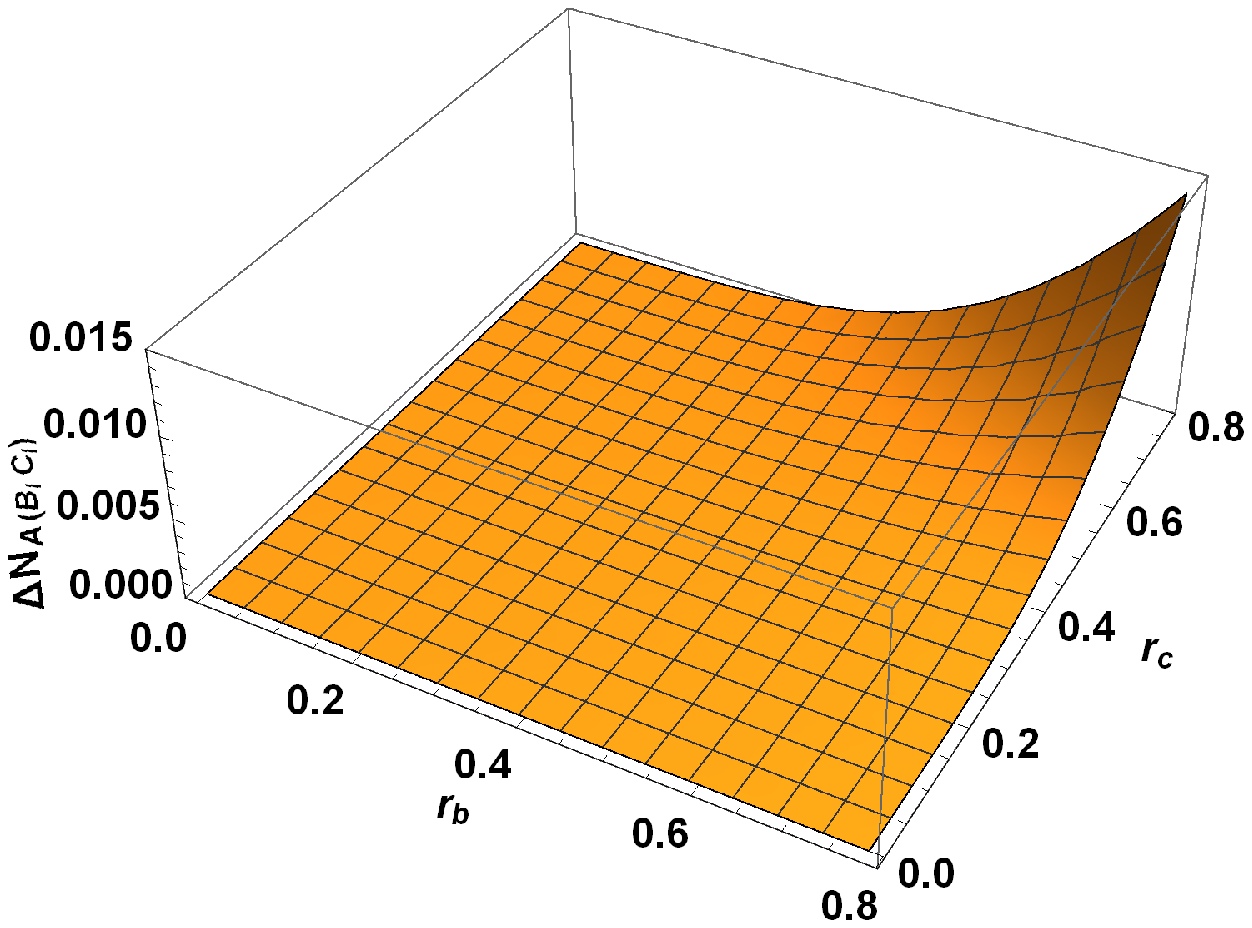}\hfil%
\caption{\label{deltaNAB1C1}(Color online) A plot of $\Delta\mathcal{N}_{A(B_IC_I)}$  as a function of acceleration parameters $r_b$ and $r_c$.}
\end{figure}

\begin{figure}[htbp]
\includegraphics[width=6cm]{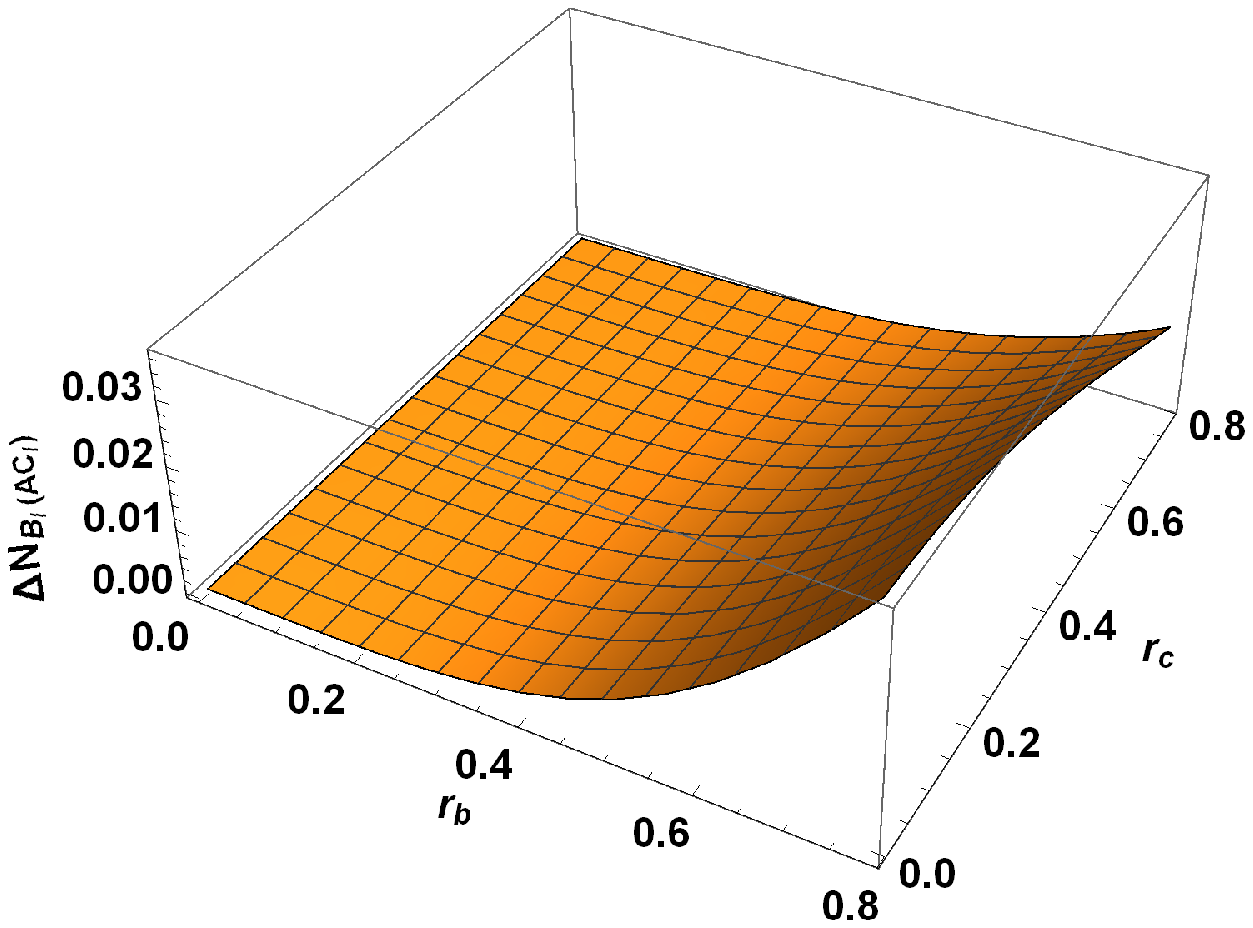}\hfill%
\caption{\label{deltaNB1AC1}(Color online) A plot of $\Delta\mathcal{N}_{B_I(AC_I)}$  as function of acceleration parameters $r_b$ and $r_c$.}
\end{figure}

\begin{figure}[htbp]
\includegraphics[width=6cm]{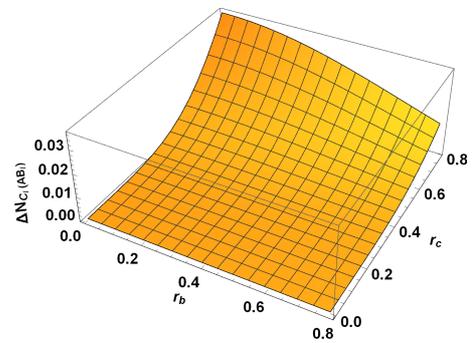}\hfil%
\caption{\label{deltaNC1AB1}(Color online) A plot of $\Delta\mathcal{N}_{C_I(AB_I)}$  as function of acceleration parameters $r_b$ and $r_c$.}
\end{figure}

\begin{figure}[htbp]
\includegraphics[width=6cm]{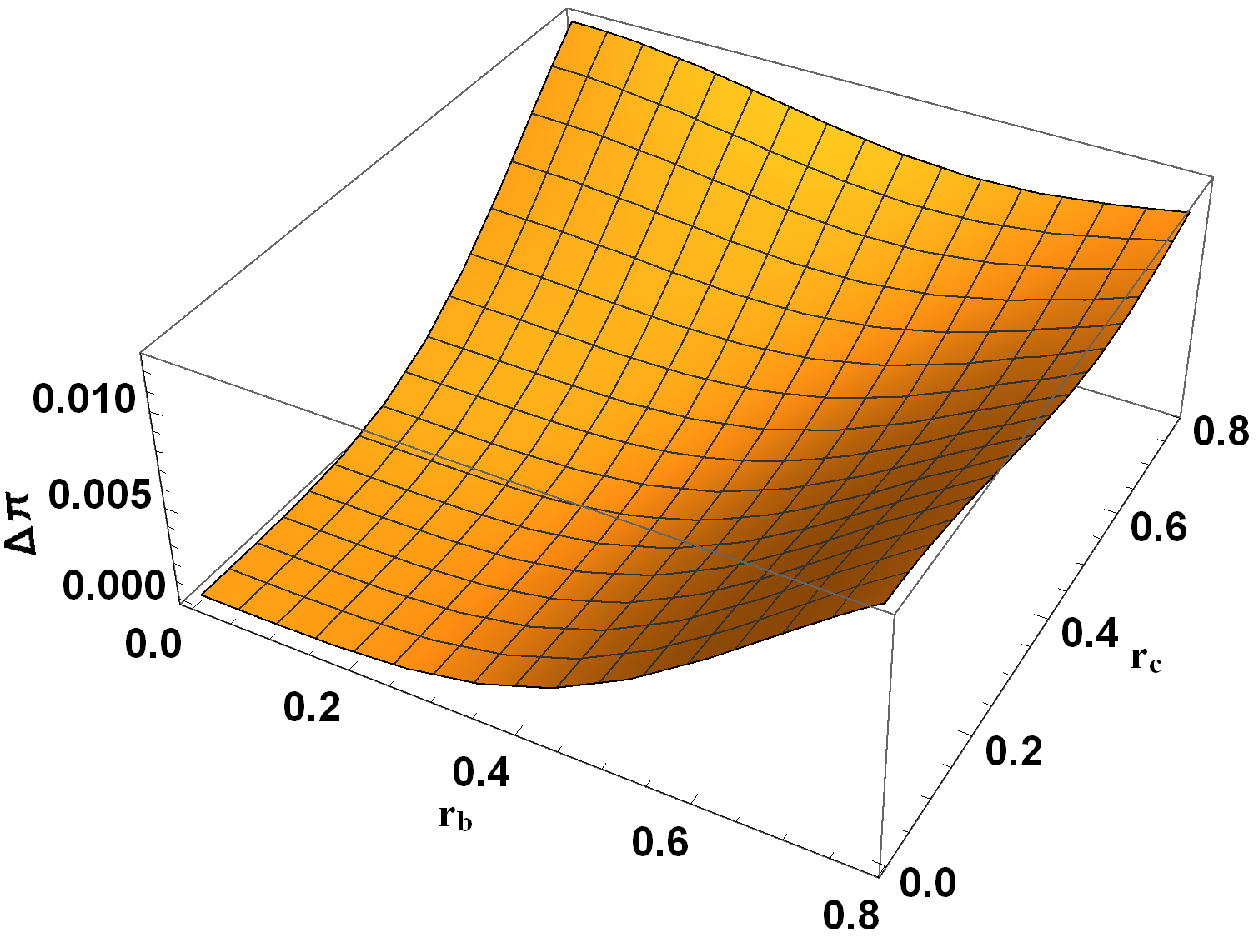}\hfil%
\caption{\label{deltap_g}(Color online) A plot of $\Delta\pi_{A(B_IC_I)}$  as function of acceleration parameters $r_b$ and $r_c$.}
\end{figure}

Let us begin with the wave-function  of this system. After making some typos corrected, Eqs.(12) and (13) in \cite{PRA_83_(2011)_022314} read
\begin{eqnarray}
 |\Phi\rangle_{AB_IB_{II}C_IC_{II}}&=& \frac{1}{\sqrt{2}}\left[\cos r_b\cos r_c|0_A0_{B_I}0_{B_{II}}0_{C_I}0_{C_{II}}\rangle \right.\nonumber\\
 &~~~~+&\cos r_b\sin r_c|0_A 0_{B_I}0_{B_{II}}1_{C_I}1_{C_{II}}\rangle  \nonumber  \\
 &~~~~+& \sin r_b \cos r_c|0_A 1_{B_I}1_{B_{II}}0_{C_I}0_{C_{II}}\rangle  \nonumber \\
 &~~~~+&\sin r_b\sin r_c|0_A 1_{B_I}1_{B_{II}}1_{C_I} 1_{C_{II}}\rangle  \nonumber \\
 &~~~~+& \left.|1_A 1_{B_I} 0_{B_{II}} 1_{C_I} 0_{C_{II}}\rangle \right], \end{eqnarray}
\begin{eqnarray}\label{rhoABICI}
\rho_{AB_IC_I}&=& \frac{1}{2}\left[\cos^2 r_b\cos^2 r_c|0_A 0_{B_I} 0_{C_I}\rangle
 \langle 0_A 0_{B_I} 0_{C_I}|\right.\nonumber\\
 &~~~~+&\cos^2 r_b\sin^2 r_c|0_A 0_{B_I}1_{C_I}\rangle\langle 0_A 0_{B_I} 1_{C_I}|  \nonumber  \\
 &~~~~+& \sin^2 r_b\cos^2 r_c|0_A 1_{B_I} 0_{C_I}\rangle
 \langle 0_A 1_{B_I} 0_{C_I}|  \nonumber \\
 &~~~~+& \sin^2 r_b\sin^2 r_c|0_A 1_{B_I} 1_{C_I}\rangle
 \langle 0_A 1_{B_I} 1_{C_I}|  \nonumber \\
 &~~~~+&\cos r_b\cos r_c(|1_A 1_{B_I} 1_{C_I}\rangle
 \langle 0_A 0_{B_I} 0_{C_I}|\nonumber\\
 &~~~~+&|0_A 0_{B_I} 0_{C_I}\rangle
 \langle 1_A 1_{B_I} 1_{C_I}|) \nonumber \\
 &~~~~+&\left.|1_A 1_{B_I} 1_{C_I}\rangle \langle 1_A 1_{B_I} 1_{C_I}|\right], \end{eqnarray}
from which we can partially transpose subsystems $A$, $B_I$ and $C_I$  of Eq.(\ref{rhoABICI}), respectively, and obtain
\begin{subequations}
\begin{eqnarray}\label{rhoABICITA}
\rho_{AB_IC_I}^{T_A}&=& \frac{1}{2}\left[\cos^2 r_b\cos^2 r_c|0_A 0_{B_I} 0_{C_I}\rangle
 \langle 0_A 0_{B_I} 0_{C_I}|\right.\nonumber\\
 &~~~~+&\cos^2 r_b\sin^2 r_c|0_A 0_{B_I}1_{C_I}\rangle\langle 0_A 0_{B_I} 1_{C_I}|  \nonumber  \\
 &~~~~+&\sin^2 r_b\cos^2 r_c|0_A 1_{B_I} 0_{C_I}\rangle
 \langle 0_A 1_{B_I} 0_{C_I}|  \nonumber \\
 &~~~~+& \sin^2 r_b\sin^2 r_c|0_A 1_{B_I} 1_{C_I}\rangle
 \langle 0_A 1_{B_I} 1_{C_I}|  \nonumber \\
 &~~~~+&\cos r_b\cos r_c|0_A 1_{B_I} 1_{C_I}\rangle
 \langle 1_A 0_{B_I} 0_{C_I}|  \nonumber \\
 &~~~~+&\cos r_b\cos r_c|1_A 0_{B_I} 0_{C_I}\rangle
 \langle 0_A 1_{B_I} 1_{C_I}| \nonumber \\
 &~~~~+&\left.|1_A 1_{B_I} 1_{C_I}\rangle \langle 1_A 1_{B_I} 1_{C_I}|\right].
 \end{eqnarray}
 \begin{eqnarray}\label{rhoABICITBI}
\rho_{A B_I C_I}^{T_{B_I}}&=&\frac{1}{2}\left [\cos^2 r_b \cos^2 r_c|0_A 0_{B_I} 0_{C_I}\rangle\langle 0_A 0_{B_I} 0_{C_I} |\right. \nonumber\\
 &~~~~+& \cos ^2 r_b\sin^2 r_c|0_A 0_{B_I} 1_{C_I}\rangle \langle 0_A 0_{B_I} 1_{C_I}| \nonumber  \\
 &~~~~+& \sin^2 r_b \cos ^2 r_c |0_A 1_{B_I} 0_{C_I}\rangle \langle 0_A 1_{B_I} 0_{CI}|\nonumber \\
 &~~~~+& \sin ^2 r_b\sin^2 r_c |0_A 1_{B_I} 1_{C_I}\rangle \langle 0_A 1_{B_I}1_{C_I}| \nonumber \\
 &~~~~+& \cos r_b\cos r_c |0_A 1_{B_I}0_{C_I}\rangle \langle 1_A 0_{B_I} 1_{CI}| \nonumber \\
 &~~~~+& \cos r_b\cos r_c|1_A 0_{B_I} 1_{C_I}\rangle \langle 0_A 1_{B_I} 0_{C_I}|\nonumber \\
 &~~~~+& \left.|1_A 1_{B_I} 1_{C_I}\rangle \langle 1_A 1_{B_I} 1_{C_I}|\right].
 \end{eqnarray}
\begin{eqnarray}\label{rhoABICITCI}
\rho_{A B_I C_I}^{T_{C_I}}&=& \frac{1}{2}\left[\cos^2 r_b\cos^2 r_c|0_A 0_{B_I} 0_{C_I}\rangle
\langle 0_A 0_{B_I} 0_{C_I}|\right.
\nonumber\\
 &~~~~+& \cos^2 r_b\sin^2 r_c|0_A 0_{B_I} 1_{C_I}\rangle\langle 0_A 0_{B_I} 1_{C_I}| \nonumber  \\
 &~~~~+& \sin^2r_b\cos^2 r_c|0_A 1_{B_I} 0_{C_I}\rangle \langle 0_A 1_{B_I} 0_{C_I}|\nonumber \\
 &~~~~+& \sin^2 r_b\sin^2 r_c |0_A 1_{B_I} 1_{C_I}\rangle \langle 0_A 1_{B_I} 1_{C_I}|\nonumber \\
 &~~~~+& \cos r_b\cos r_c|1_A 1_{B_I} 0_{C_I}\rangle \langle 0_A 0_{B_I} 1_{C_I}|\nonumber \\
 &~~~~+& \cos r_b\cos r_c |0_A 0_{B_I} 1_{C_I}\rangle \langle 1_A 1_{B_I} 0_{C_I}|\nonumber \\
 &~~~~+& |1_A 1_{B_I} 1_{C_I}\rangle \langle 1_A 1_{B_I} 1_{C_I}|.
 \end{eqnarray}
 \end{subequations}
The $ \rho_{A B_I C_I}^{T_A}, \rho_{A B_I C_I}^{T_{B_I}}$ and $\rho_{A B_I C_I}^{T_{C_I}}$ are given in the following matrix form
\begin{widetext}
\begin{subequations}
\begin{equation}\label{rhoABICITAm}
    \rho_{A B_I C_I}^{T_A}=\frac{1}{2}\left(
 \begin{array}{cccccccc}
% \begin{smallmatrix}
 \cos ^2 r_b\cos^2r_c & 0 & 0 & 0 & 0 & 0 & 0 & 0 \\
 0 & \cos^2 r_b \sin^2 r_c & 0 & 0 & 0 & 0 & 0 & 0 \\
 0 & 0 &  \sin^2 r_b \cos^2 r_c & 0 & 0 & 0 & 0 & 0 \\
 0 & 0 & 0 &  \sin^2 r_b \sin^2 r_c & \cos r_b \cos r_c & 0 & 0 & 0 \\
 0 & 0 & 0 &  \cos r_b \cos r_c & 0 & 0 & 0 & 0 \\
 0 & 0 & 0 & 0 & 0 & 0 & 0 & 0 \\
 0 & 0 & 0 & 0 & 0 & 0 & 0 & 0 \\
 0 & 0 & 0 & 0 & 0 & 0 & 0 & 1 \\
% \end{smallmatrix}
\end{array}
\right),
\end{equation}
\begin{equation}\label{rhoABICITBIm}
    \rho_{A B_I C_I}^{T_{B_I}}=\frac{1}{2}
 \left(
\begin{array}{cccccccc}
  \cos ^2 r_b \cos^2 r_c & 0 & 0 & 0 & 0 & 0 & 0 & 0 \\
 0 & \cos^2 r_b \sin^2 r_c & 0 & 0 & 0 & 0 & 0 & 0 \\
 0 & 0 & \sin^2 r_b \cos^2 r_c  & 0 & 0 & \cos r_b \cos r_c & 0 & 0 \\
 0 & 0 & 0 & \sin ^2 r_b \sin ^2 r_c & 0 & 0 & 0 & 0 \\
 0 & 0 & 0 & 0 & 0 & 0 & 0 & 0 \\
 0 & 0 &  \cos r_b \cos r_c & 0 & 0 & 0 & 0 & 0 \\
 0 & 0 & 0 & 0 & 0 & 0 & 0 & 0 \\
 0 & 0 & 0 & 0 & 0 & 0 & 0 & 1 \\
\end{array}
\right),
% \end{smallmatrix}
\end{equation}
\begin{equation}\label{rhoABICITCIm}
    \rho_{A B_I C_I}^{T_{C_I}}=\frac{1}{2}\left(
\begin{array}{cccccccc}
 \cos^2 r_b\cos^2 r_c & 0 & 0 & 0 & 0 & 0 & 0 & 0 \\
 0 & \cos^2 r_b\sin^2 r_c & 0 & 0 & 0 & 0 & \cos r_b \cos r_c & 0 \\
 0 & 0 & \sin^2 r_b \cos^2 r_c  & 0 & 0 & 0 & 0 & 0 \\
 0 & 0 & 0 & \sin^2 r_b \sin^2 r_c & 0 & 0 & 0 & 0 \\
 0 & 0 & 0 & 0 & 0 & 0 & 0 & 0 \\
 0 & 0 & 0 & 0 & 0 & 0 & 0 & 0 \\
 0 & \cos r_b \cos r_c & 0 & 0 & 0 & 0 & 0 & 0 \\
 0 & 0 & 0 & 0 & 0 & 0 & 0 & 1 \\
\end{array}
\right).
% \end{smallmatrix}
\end{equation}
\end{subequations}
\end{widetext}
Based on $\rho_{A B_I C_I}^{T_A}, \rho_{A B_I C_I}^{T_{B_I}}$ and $\rho_{A B_I C_I}^{T_{C_I}}$, we can calculate $\pi$-tangle as \cite{PRA_75_(2007)_062308}
\begin{equation}\label{pi_d}
   \pi_{\alpha\beta\gamma}=\frac{1}{3}(\pi_\alpha+\pi_\beta+\pi_\alpha\gamma),
\end{equation}
where
\begin{subequations}
\begin{equation} \pi_\alpha= \mathcal{N}_{\alpha(\beta\gamma)}^2-\mathcal{N}_{\alpha\beta}^2 -\mathcal{N}_{\alpha\gamma}^2,
\end{equation}
 \begin{equation}\pi_\beta = \mathcal{N}_{\beta(\alpha\gamma)}^2-\mathcal{N}_{\beta\alpha}^2 -\mathcal{N}_{\beta\gamma}^2, \end{equation}
 \begin{equation} \pi_\gamma = \mathcal{N}_{\gamma(\alpha\beta)}^2-\mathcal{N}_{\gamma\alpha}^2 -\mathcal{N}_{\gamma\beta}^2.
 \end{equation}
\end{subequations}
The so-called "two-tangle" $\mathcal{N}_{AB}=\|\rho_{AB}^{T_A}\|-1$ represents the negativity of the mixed state $\rho_{AB}$, while the "one-tangle" is given by $\mathcal{N}_{A(BC)}=\|\rho_{ABC}^{T_A}\|-1$, and $\|A\|=\mbox{tr}(\sqrt{AA^\dag})$ is the trace norm of matrix $A$, i.e. the
sum of the singular values of $A$ \cite{MA}. Alternatively, $\|A\|-1$ is equal to two times of the sum of  absolute values of negative eigenvalues of $A$.
\begin{subequations}
\begin{eqnarray}\label{NABICI2}
  \mathcal{N}_{A(B_IC_I)}&=&  \frac{1}{2}\Big(\sqrt{\sin^4 r_b\sin^4 r_c+4 \cos^2 r_b \cos^2 r_c}\nonumber\\
  &~~~~~-&\sin^2 r_b \sin ^2 r_c\Big),
\end{eqnarray}
\begin{eqnarray}\label{NBIACI2}
  \mathcal{N}_{B_I(AC_I)}&=& \frac{1}{2} \cos r_c \Big(\sqrt{\sin^4 r_b \cos^2 r_c+4 \cos^2 r_b}\nonumber\\
  &~~~~~-& \sin ^2 r_b \cos r_c\Big),
\end{eqnarray}
\begin{eqnarray}\label{NCIABI2}
  \mathcal{N}_{C_I(AB_I)}&=&\frac{1}{2} \cos r_b\Big(\sqrt{\cos^2 r_b\sin^4 r_c+4 \cos^2 r_c}\nonumber\\
  &~~~~~-&\cos r_b \sin^2 r_c\Big).
\end{eqnarray}
\end{subequations}
 Using our formulae of one-tangle, we find when Bob and Charlie move with infinite acceleration ($r_b=r_c=\pi/4$), $\mathcal{N}_{A(B_IC_I)}=\mathcal{N}_{B_I(AC_I)}=\mathcal{N}_{C_I(AB_I)}=(\sqrt{17}-1)/8$ instead of $(1-\sqrt{5})/8$. The first and third formulae of Eq.(19) in \cite{PRA_83_(2011)_022314} are correct, but the second one should be
\begin{equation}\label{19}
\mathcal{N}_{C_I(AB_I)}=\frac{1}{2}\left (\sqrt{4 \cos^2 r_c+\sin^4 r_c}-\sin^2 r_c\right).
\end{equation}
Because $\mathcal{N}_{AB_I}=\mathcal{N}_{AC_I}=\mathcal{N}_{B_IA}=\mathcal{N}_{B_IC_I}=\mathcal{N}_{C_IA}=\mathcal{N}_{C_IB_I}=0$,
equation (18) of \cite{PRA_83_(2011)_022314} is still valid
\begin{eqnarray}\label{pp}
  \pi_{AB_IC_I} &=&\frac{1}{3}( \pi_A+\pi_{B_I}+\pi_{C_I}) \nonumber\\
   &=& \frac{1}{3}\left[\mathcal{N}_{A(B_IC_I)}^2+\mathcal{N}_{B_I(AC_I)}^2+\mathcal{N}_{C_I(AB_I)}^2\right].
\end{eqnarray}
To see the difference between our results and ones in \cite{PRA_83_(2011)_022314}, we plot in Figs.\ref{deltaNAB1C1}, \ref{deltaNB1AC1} and \ref{deltaNC1AB1}, respectively the differences of one-tangles calculated using Eqs.(\ref{NABICI}), (\ref{NBIACI}), (\ref{NCIABI}) and  ones derived from   Eqs.(\ref{NABICI2}), (\ref{NBIACI2}), (\ref{NCIABI2}).
In these figures,  $\Delta\mathcal{N}_{A(B_IC_I)}$, $\Delta\mathcal{N}_{B_I(AC_I)}$  and $\Delta\mathcal{N}_{C_I(AB_I)}$ denote the differences of right sides of Eqs.(\ref{NABICI}), (\ref{NABICI2}), Eqs.(\ref{NBIACI}), (\ref{NBIACI2}) as well as Eqs.(\ref{NCIABI}), (\ref{NCIABI2}), respectively. We also plot the difference of $\pi$-tangle calculated based on Eqs.(\ref{NABICI}), (\ref{NBIACI}), (\ref{NCIABI}) and  one obtained from   Eqs.(\ref{NABICI2}), (\ref{NBIACI2}), (\ref{NCIABI2}) in Fig.\ref{deltap_g}.
To further investigate the difference between our results and ones in \cite{PRA_83_(2011)_022314} as illustrated in Fig.\ref{deltap_g}, we expand $\Delta\pi_{A(B_IC_I)}$ as series of $r_b$ and $r_c$ to fifth order,
\begin{equation}\label{deltap}
    \Delta\pi_{A(B_IC_I)}\approx\frac{1}{12} (r_b^4+r_c^4)-\frac{1}{6}r_b^2 r_c^2(r_b^2+r_c^2)+\frac{13}{36}r_b^4 r_c^4.
\end{equation}
Figures \ref{deltaNAB1C1}, \ref{deltaNB1AC1}, \ref{deltaNC1AB1}, \ref{deltap_g} and Eq.(\ref{deltap}) show that the differences of one-tangle and $\pi$-tangle calculated according to Eqs.(\ref{NABICI}), (\ref{NBIACI}), (\ref{NCIABI}) and  Eqs.(\ref{NABICI2}), (\ref{NBIACI2}), (\ref{NCIABI2}),
respectively, become more obvious for larger  $r_b$ and $r_c$.
%\vspace{2cm}
In conclusion, we have corrected three key formulae about one-tangle of multipartite entanglement of fermionic systems in noninertial frames in Phys. Rev. A 83, 022314(2011) and illustrated their differences explicitly.


\begin{thebibliography}{99}
\bibitem {PRA_74_(206)_032326} P. M. Alsing, I. Fuentes-Schuller, R. B. Mann, and T. E. Tessier,
Phys. Rev. A \textbf{74} (2006) 032326.
\bibitem {PRA_83_(2011)_022314} J. Wang and J. Jing, Phys. Rev. A \textbf{83} (2011) 022314.
\bibitem {PRA_75_(2007)_062308}Y. C. Ou and H. Fan, Phys. Rev. A \textbf{75} (2007) 062308.
\bibitem {MA}R. A. Horn and C. R. Johnson, Matrix Analysis (Cambridge University Press, New York, 1985), p.205, 415, 441.
\end{thebibliography}
\end{document}